\documentclass[twocolumn]{aastex631}

\usepackage{xspace}
\newcommand{\swift}{{\it Swift}\xspace}
\newcommand{\nicer}{\textit{NICER}\xspace}

\shorttitle{A superflare on HD 251108}
\shortauthors{G\"unther et al.}

\graphicspath{{./}{figures/}}

\usepackage{xcolor}

\begin{document}

\title{A long-duration superflare on the K giant HD 251108}

%% The new \altaffiliation can be used to indicate some secondary information
%% such as fellowships. This command produces a non-numeric footnote that is
%% set away from the numeric \affiliation footnotes.  NOTE that if an
%% \altaffiliation command is used it must come BEFORE the \affiliation call,
%% right after the \author command, in order to place the footnotes in
%% the proper location.

\correspondingauthor{Hans Moritz G{\"u}nther}
\email{hgunther@mit.edu}

% OK for submission: Scott, Rakesh, Carl

\author[0000-0003-4243-2840]{Hans Moritz G{\"u}nther}
\affiliation{MIT Kavli Institute for Astrophysics and Space Research, 77 Massachusetts Avenue, Cambridge, MA 02139, USA}

\author[0000-0003-1386-7861]{Dheeraj Pasham}
\affiliation{MIT Kavli Institute for Astrophysics and Space Research, 77 Massachusetts Avenue, Cambridge, MA 02139, USA}

\author[0000-0001-5976-4644]{Alexander Binks}
\affil{Institut f\"ur Astronomie und Astrophysik, Eberhard-Karls Universit\"at T\"ubingen, Sand 1, 72076 T\"ubingen, Germany}

\author[0000-0002-4203-4773]{Stefan Czesla}
\affil{Thüringer Landessternwarte Tautenburg, Sternwarte 5, 07778 Tautenburg, Germany}

\author[0000-0003-1244-3100]{Teruaki Enoto}
\affil{Department of Physics, Kyoto University, Sakyo, Kyoto 606-8502, Japan}
\affiliation{RIKEN Cluster for Pioneering Research, 2-1 Hirosawa, Wako, Saitama 351-0198, Japan}

\author[0000-0002-9113-7162]{Michael Fausnaugh}
\affil{Texas Tech University, Department of Physics \& Astronomy, Lubbock, TX 79409, USA}

\author[0000-0003-0125-8700]{Franz-Josef Hambsch}
\affiliation{Vereniging Voor Sterrenkunde (VVS), Oostmeers 122 C, 8000 Brugge, Belgium}
\affiliation{AAVSO, 185 Alewife Brook Parkway, Suite 410, Cambridge, MA 02138, USA}
\affiliation{Groupe Européen d’Observations Stellaires (GEOS), 23 Parc de Levesville, 28300 Bailleau l’Evêque, France }
\affiliation{Bundesdeutsche Arbeitsgemeinschaft für Veränderliche Sterne e.V. (BAV), Munsterdamm 90, 12169 Berlin, Germany}

\author[0000-0003-3085-304X]{Shun Inoue}
\affil{Department of Physics, Kyoto University, Sakyo, Kyoto 606-8502, Japan}

\author[0000-0003-0332-0811]{Hiroyuki Maehara}
\affil{Okayama Branch Office, Subaru Telescope, National Astronomical Observatory of Japan, NINS, Kamogata, Asakuchi, Okayama 719-0232, Japan}

\author[0000-0002-0412-0849]{Yuta Notsu}
\affil{Laboratory for Atmospheric and Space Physics, University of Colorado Boulder, 3665 Discovery Drive, Boulder, CO 80303, USA}
\affil{National Solar Observatory, 3665 Discovery Drive, Boulder, CO 80303, USA}

\author[0000-0002-4939-8940]{Jan Robrade}
\affiliation{Hamburger Sternwarte, University of Hamburg, Gojenbergsweg 112, 21029 Hamburg, Germany}

\author[0000-0003-2554-9916]{J. H. M. M. Schmitt}
\affiliation{Hamburger Sternwarte, University of Hamburg, Gojenbergsweg 112, 21029 Hamburg, Germany}

\author[0000-0002-5094-2245]{P. C. Schneider}
\affiliation{Hamburger Sternwarte, University of Hamburg, Gojenbergsweg 112, 21029 Hamburg, Germany}

%% AASTeX 6.31 has the new \collaboration and \nocollaboration commands to
%% provide the collaboration status of a group of authors. These commands
%% can be used either before or after the list of corresponding authors. The
%% argument for \collaboration is the collaboration identifier. Authors are
%% encouraged to surround collaboration identifiers with ()s. The
%% \nocollaboration command takes no argument and exists to indicate that
%% the nearby authors are not part of surrounding collaborations.

%% Mark off the abstract in the ``abstract'' environment.
\begin{abstract}
Many giant stars are magnetically active, which causes rotational variability, chromospheric emission lines, and X-ray emission. Large outbursts in these emission features can set limits on the magnetic field strength and thus constrain the mechanism of the underlying dynamo. HD~251108 is a Li-rich active K-type giant. We find a rotational period of 21.3~d with color changes and additional long-term photometric variability. Both can be explained with very stable stellar spots.
We followed the decay phase of a superflare for 28 days with NICER and from the ground. We track the flare decay in unprecedented detail in several coronal temperature components. With a peak flux around $10^{34}$~erg~s$^{-1}$ (0.5-4.0~keV) and an exponential decay time of 2.2~days in the early decay phase, this is one of the strongest flares ever observed; yet it follows trends established from samples of smaller flares, for example for the relations between H$\alpha$ and X-ray flux, indicating that the physical process that powers the flare emission is consistent over a large range of flare energies. We estimate a  flare loop length about 2-4 times the stellar radius. No evidence is seen for abundance changes during the flare.

\end{abstract}

\section{Introduction} \label{sec:intro}
Flares are an obvious sign of magnetic activity on the Sun and other late-type (spectral type M-F) stars, which have convective outer envelopes. Differential rotation and turbulence give rise to dynamo action including surface magnetic fields \citep[review by][]{2017LRSP...14....4B}.
%This is well-studied on the Sun, where magnetic fields and chromospheric and coronal activity are temporally and spatially resolved.
%A closed magnetic field line can get twisted because its foot points on the stellar surface move around due to differential rotation and plasma motion.
When magnetic field lines reconnect, they accelerate non-thermal electrons towards the stellar surface, where they heat up material which expands into the magnetic loop and drives a shock wave \citep{Parker_1988}.
%The electrons can be seen in synchrotron radio emission and hard X-rays ($> 10$~keV), followed a few minutes later by soft X-rays ($<10$~keV) from the thermal plasma heated by the shock \citep{Neupert_1968}.
The shock wave travels along the magnetic loop and increases the density and temperature of the confined plasma, which will cool through soft X-ray radiation. The density, temperature and decay time of the flare are related to the geometric properties of the loop \citep[e.g.][]{Reale_2004}. On the Sun, flares often occur in arcades of several closely spaced magnetic field lines at the same time, but fortunately, a single loop model can still provide a good description if  one of the loops clearly dominates the emission \citep{Reale_2004} or most of the loops in the arcade light up at the same time \citep{Getman_2011}. More luminous flares reach higher temperatures and last longer in the Sun and other dwarf stars \citep[see review by][]{2009A&ARv..17..309G} and total energy output and flare duration also follow a powerlaw in RS~CVn stars \citep{2023MNRAS.518..900K}.

RS~CVn stars are binaries composed of a cool giant or sub-giant and a smaller (sub-giant or main-sequence) companion. Long-lasting and large scale X-ray flares have been observed on a number of cool giants, e.g.,\ in {\it ROSAT} observations of a 1.5~d duration flare on the RS CVn-type binary \object{HU Virginis} \citep{1997A&A...328..565E}, a 9~d duration event on \object{CF Tuc} \citep{1996A&A...311..211K} and more recently a $>1.3$~d long flare observed on \object{SZ Psc} with \emph{Swift} \citep{2023MNRAS.518..900K} and a flare lasting almost a week on \object{UX Ari} with \nicer \citep{2024ApJ...965..135K}; see \citet{2023MNRAS.518..900K} for a compilation of large flares on RS~CVn stars. In contrast, even large flares on main-sequence stars typically last only a few hours. RS~CVn stars in close orbits can be tidally locked, leading to rotation periods of just a few days and short rotation periods provide more energy for stellar activity. Less is known about X-ray flares on single giants but \citet{2022A&A...668A.101O} analyze thousands of optical flares on giant stars in the Kepler field, many of which appear to be single. They find that the flare shapes are similar to dwarf stars, but they have a longer duration and higher energy output and that flare frequency and energy distributions are also similar to dwarf stars.

In this paper, we present new data on one of the largest flares ever observed. First, we determine properties of  HD~251108, which establish it as a magnetically active giant star. We estimate temperature, mass, and other stellar properties and take a look at the optical lightcurve in different bands (section~\ref{sect:target}). In the second part (section~\ref{sect:flare}), we describe new observations of a new long-duration X-ray flare seen on HD~251108 discovered by the Lobster Eye Imager for Astronomy \citep[LEIA,][]{2023arXiv230514895L} and announced by \citet{2022ATel15748....1L}. We present follow-up observations with the Neutron Star Interior Composition Explorer \citep[NICER,][]{2017NatAs...1..895G} and other telescopes. Thanks to the superior effective area of \nicer, we obtain more detailed X-ray data on the cooling phase of the flare than any previous observation of giant flares.

% Maybe is sentence can be commented out:
%First, we summarize the most important properties of HD~251108, a K-type giant, in section~\ref{sect:target}. In section~\ref{sect:data}, we describe observations and data reduction for the data used in this work. We derive results in the remainder of section~\ref{sect:results}. We discuss the results in section~\ref{sect:discussion} and summarize our findings in section~\ref{sect:summary}.

\section{Properties of HD 251108}
\label{sect:target}

HD~251108 is an evolved, cool star, located above the main sequence in a color-magnitude diagram. GAIA \citep{2016A&A...595A...1G,2022arXiv220800211G} counterpart 3342754943193083392 has a distance of $505\pm5$~pc \citep{2021AJ....161..147B} with a proper motion of a few milliarcsecond per year \citep{2020yCat.1350....0G}.
Its GAIA magnitude is $m_G=9.600\pm0.006$, about 4 magnitudes brighter than any other source in a 30\arcsec{} radius. Thus, contamination of optical spectroscopy is unlikely, and we treat any contamination of photometry as negligible.

The Tess Input Catalog v.8.2 \citep{2021arXiv210804778P} lists $T_\mathrm{eff}=4460\pm130$~K, a giant luminosity class, a radius around $13 R_\sun$, and an $E(B-V)=0.1$~mag reddening and \citet{2012A&A...537A..91X} determine $T_\mathrm{eff}=4339$~K from a spectroscopic line ratio. As we will show below, the star is time-variable in the optical, and in section~\ref{sect:color}, we find $T_\mathrm{eff}=4450\pm50$~K in the bright state.
Placing the photometry on evolutionary tracks, we find $M_*\approx1\;M_\sun$ \citep[see, e.g., Fig.~3 in ][]{2018MNRAS.480.2137S}.
The stellar radius is $R_\star=6-8\;R_\sun$, with estimates using mass and surface gravity from GAIA or magnitude, distance, and temperature giving a similar result. Our estimate for $R$ is thus about 30-40\% smaller than the vlaue in the TIC even after accounting for the improved GAIA distance available to us. A forthcoming paper analyzing a time series of optical spectroscopy \citep{Fuhrmeister} will provide a more detailed analysis of the stellar properties.
% M_G = 9.6 - 5 * (np.log10(505) - 1) = 1.08

% g band from ASAS-SN - bright star
% M_g = 10.5 - 5 * (np.log10(505) - 1) = 2.0
% Hanbsch, bright statre
% M_V = 10 - 5 * (np.log10(505) - 1) = 1.5
% Mamajek table is for dwarfs, but BC should be similar since it mostly
% depends on T_eff
% BC_V = -0.63
% M_bol = 1.5 - 0.63 = 0.87
% M_bol_target = 1.5 - 0.63 = 0.87
% M_bol_sun = 4.74
% L_target / L_sun = 10^((4.74 - 0.87) / -2.5) = 5.5

% 2017MNRAS.471..722H lists  +0.44 ± 0.01 (in G) for RC with mean intrinsic dispersion is ∼0.17 ± 0.03 mag across all bands

% https://ui.adsabs.harvard.edu/abs/1996ApJ...469..355F/abstract also has BC

\begin{figure}
    \centering
    \includegraphics[width=0.45\textwidth]{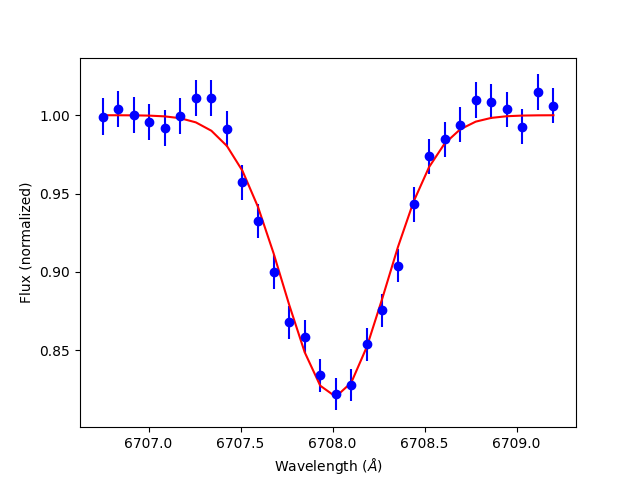}
    \caption{TIGRE optical spectrum: the presence of the Li line \deleted{(left)} and He~I in emission (not shown) indicates that HD~251108 belongs to the subgroup of Li-rich, active giants.
    \label{fig:opt_spec}}
\end{figure}

GAIA distance and flux in G or H band \citep{2003yCat.2246....0C} imply that HD~251108 is a red-clump star \citep{2017MNRAS.471..722H}.
Such stars are prominently seen in flux-limited X-ray surveys and ground-based surveys show that some red clump stars have both high Li-abundances and active chromospheres; the reason for this phenomenon may be related to the exact evolutionary phase of the star: Most Li-rich, active stars are He-core-burning red clump or red horizontal branch stars and they have a higher fraction of binaries than Li-poor red giants \citep[][]{Sneden_2022}.
We detect the Li absorption line (Figure~\ref{fig:opt_spec}, see section~\ref{sect:optspec} and \citet{Fuhrmeister} for details) with an equivalent width compatible to a measurement about 15 years earlier by \citet{2012A&A...537A..91X}. %He~{\sc i} at 5876~\AA{} is also in emission. This line is variable throughout the observations and places HD~251108 in the subgroup of Li-rich, active giants.
Thus, HD~251108 likely belongs to the small but well-known subgroup of giants that are magnetically active.

\subsection{Archival data and observations}
\label{sect:longtermphotometry}

\subsubsection{DASCH}
The longest coverage comes from the DASCH project \citep[Digital Access to a Sky Century at Harvard,][]{2012IAUS..285..243G,2013PASP..125..857T}, which digitized a diverse collection of photographic plates from the Harvard College Observatory. Data is taken with different telescopes and widely varying instrument setups. In particular, the different plates used over the years have different sensitivity that do not necessarily correspond to modern standard filter bands. The DASCH pipeline corrects the photometry to the band of a reference catalog, but in particular for objects that change color over time, this correction is not accurate.
We ignore instances where only an upper limit on HD~251108 is available, since it is often not clear how good those limits are. For photometry, we restrict the analysis to data points with no warnings from the photometric pipeline (i.e.\ \texttt{AFLAGS ==0}). The earliest remaining datapoint is from 1889, with most observations in the periods 1900-1955 and 1975-1990.

\subsubsection{ASAS-3}
Data for HD~251108 is available from the All Sky Automated Survey \citep[ASAS-3,][]{2002AcA....52..397P} for multiple observing seasons between 2003 and 2008. ASAS-3 observed in the $V$ band and contains data for different extraction regions in aperture photometry with a radius between 2 and 6 pixels. For a bright star like our target, differences are small. We chose the extraction with a radius of 4 pixels.

\subsubsection{KWS}
The Kamogata/Kiso/Kyoto Wide-field Survey \citep[KWS,][]{oai:jaxa.repo.nii.ac.jp:00001963} provides $V$ and $I$ band data from about 2014 on. We remove data points with uncertainties $>0.1$~mag and average multiple observations taken in the same night (usually within a few minutes).

\subsubsection{ASAS-SN}
\label{sect:asassn}
Photometric data also comes from ASAS-SN \citep{2014ApJ...788...48S,2017PASP..129j4502K} and we specifically used the ASAS-SN aperture photometry pipeline. $V$ band photometry is likely saturated for a source as bright as HD~251108, so we restrict this analysis to use only $g$ band data. We retrieved 2644 photometric datapoints. While the uncertainties for most datapoints are about 0.005~mag, a tail in the distribution to much larger uncertainties exists. We only make use of data with uncertainties $<0.007$~mag, which removed 199 datapoints.

\subsubsection{AAVSO}
We retrieved observations from the AAVSO database \citep{AAVSO}. Multiple observers began monitoring HD~251108 after the flare discovery, but they use different instruments and standard stars which leads to visible offsets in the observed magnitudes. For the analysis of the color evolutions, we thus restrict our analysis to data from a single observer (AAVSO observer code: HMB), who obtained photometry in $BVRI$ almost nightly for 175 days, collecting two exposures per band using the Remote Observatory Atacama Desert  \citep[ROAD,][]{2012JAVSO..40.1003H} in Astrodon $B$, $V$, $R$, and $I$ bands. %It consists of a 40-cm f /6.8 Optimized DallKirkham and uses a Finger Lakes Instruments camera with a 4kx4k array with pixels of 9~µm in size. Data are reduced using a custom pipeline and then published on the AAVSO website.
We average both exposures in a single band in each night. Exposures in different bands are taken consecutively, and we assume that the star does not vary on timescales of a few minutes.

\subsubsection{TESS}
HD~251108 is also covered in five sectors by the Transiting Exoplanet Survey Satellite \citep[TESS,][]{2015a_Ricker} mission. We extracted light curves using difference imaging, which gives a reliable relative light curve within each TESS sector. Global offsets in the TESS magnitude between sectors are caused by uncertainties in the local background and crowding from nearby stars, due to TESS's large pixels (21$\times$21 arcseconds). The data in sector~45 (and to a lesser degree sector~44) suffers from an artifact in the image subtraction sometimes associated with brighter stars that leads to unreliable photometric measurements. Full details of the TESS data reduction and light curve extraction are in \citet{2023ApJ...956..108F}.

TESS data in Sector~06 were taken at a 30 minute cadence, and TESS data in Sectors~33, 43, 44, and 45 were taken at a 10 minute cadence. TESS observes in a passband from 600--1000\,nm.

\subsubsection{ROSAT}
\citet{2022A&A...664A.105F} identify 2RXS J060415.1+124554 with HD 251108 with a probability of 93\%. The ROSAT source has
a total of $164\pm14$~ct. There are indications for variability between $0.15\pm0.17$ and $0.59\pm0.15$~cts~s$^{-1}$ and the spectrum shows a high hardness ratio of 0.82 \citep{2016A&A...588A.103B}, indicating a hot corona in the ROSAT data. The average ROSAT count rate corresponds to a luminosity of $\log L_\mathrm{X} = 32.2$ in erg~s$^{-1}$ extrapolated to the 0.5-4.0 keV range using the absorbing column density found for the NICER data (section~\ref{xray:spectra}) and the GAIA distance; for plasma temperatures between 0.5 and 5~keV, the count rate to flux conversion differs only by a factor of two, so the estimated $\log L_\mathrm{X}$ is valid for all reasonable coronal temperatures.

\subsubsection{XMM-Newton slew survey}
HD~251108 was observed by XMM-Newton in the slew survey on 2016-10-12 (ObsID 9308400004) for a total exposure time in the PN of about 8~s. We retrieved the pipeline-processed data from the archive and selected a large, apparently source-free region with a similar exposure time as background. A source is detected with net rate of $2.6\pm0.6$ counts~s$^{-1}$ within 8~arcsec \citep[the typical $1\sigma$ positional uncertainty of sources in the slew survey,][]{2008A&A...480..611S}.  With the same assumptions as for the ROSAT spectrum, we again find a luminosity about $\log L_\mathrm{X} = 32.2$ in erg~s$^{-1}$.

\subsubsection{eROSITA}

eROSITA \citep{2021A&A...647A...1P} is the soft X-ray instrument onboard the SRG spacecraft \citep{2021A&A...656A.132S} and performs an all-sky survey in the 0.2\,--\,10.0~keV energy range, the eROSITA all-sky survey (eRASS). The survey itself is composed of a series of half-year long individual all-sky surveys, here we use eRASS1 to eRASS4. During each half-year survey any sky-location is again observed multiple times by eROSITA in progressing and overlapping stripes with a scan rotation period of 4~h. In the eRASS image HD~251108 is a well separated X-ray source with no other relevant sources within its 5' vicinity. The positional cross-match with the optical position is excellent, given an angular separation of 0.5" and positional errors of about 0.3" (stat) and 1.0" (sys).

HD~251108 is detected in eRASS1-eRASS4 (eRASS1/DR1: 1eRASS J060415.1+124550) with an average X-ray luminosity of $\log L_\mathrm{X} = 32.2$ [erg\,s$^{-1}$] in the 0.2\,--\,2.3 keV eRASS band, corresponding to $\log L_\mathrm{X} = 32.3$ [erg\,s$^{-1}$] in the 0.5\,--\,4.0 keV band. The survey count rates, i.e.\ including correction for vignetting, PSF fraction etc., vary in the range of 4\,--\,8 cts\,s$^{-1}$ among the individual all-sky surveys, with slightly harder emission during the brighter states as deduced from median photon energies ranging between 950~eV and 975~eV. On short timescales variability at the 10\,--\,20\,\% level is seen within each eRASS, but no exceptional flaring or other larger brightness excursions are present. In total about 2250 source counts are detected, sufficient for a spectral analysis (see Sect.~\ref{sect:eRASSresults}).

\subsection{Results}

\begin{figure*}
    %\begin{interactive}{js}{lc_opt_interactive.zip}
        \centering
        \includegraphics[width=0.9\textwidth]{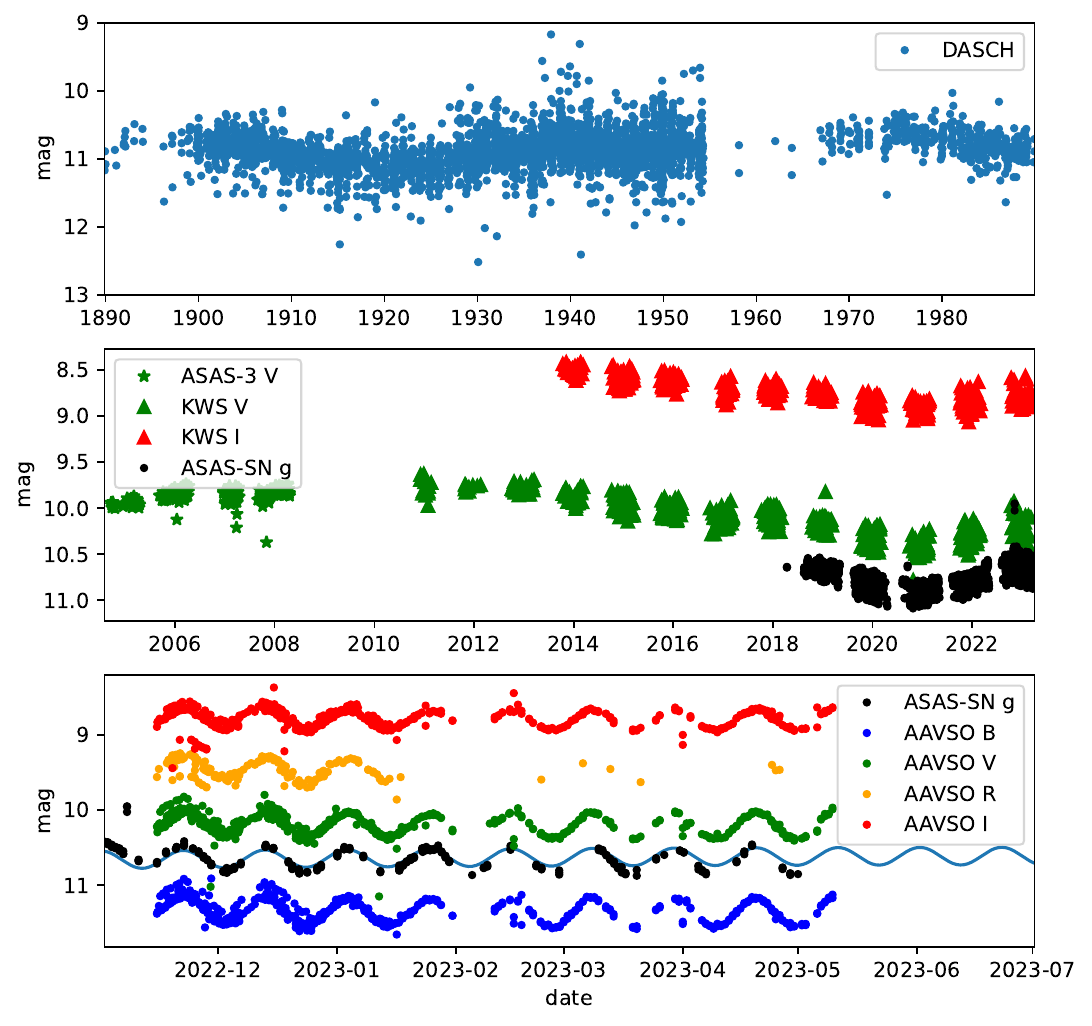}
    %\end{interactive}
    \caption{Optical light curves of HD~251108. Error bars are omitted for clarity. In all panels the blue curve shows a model of two sinusoids, fitted to the ASAS-SN data. \emph{top:} DASCH light curve. \emph{middle:} ASAS-SN light curve. The flare on 2022-11-07 is easily visible and is the brightest flare observed, a second flares is seen on 2020-09-13.  \emph{bottom:} ASAS-SN and multi-band AAVSO light curve over just a few of the short periods. The AAVSO data is taken from different observers. Inconsistent calibration and different instruments lead to systematic offsets within one band (e.g.\ $R$ band in 2022-11) between different observers, but the overall variability pattern is consistent.
    An interactive version of this figure is available in the online journal.%, where all the data are shown in a single panel and the user can zoom in and select a range of dates to see the data in more detail.
    \label{fig:optlc}}
\end{figure*}
\label{sect:optlc}
\subsubsection{Periodicity}

The ASAS-SN lightcurve (Figure~\ref{fig:optlc}, middle and bottom) can be fit well with a combination of two sinusoids. The shorter period is $21.3$d with a zero point of $58229.1$ in MJD. The amplitude is about 0.02~mag with a very low formal fit uncertainty. However, figure~\ref{fig:optlc} shows cycle-to-cycle variations in both the amplitude and the shape of the light curve (best visible in the bottom panel or in a zoom in the interactive figure in the online version). We searched for variations in the period between the different ASAS-SN observing seasons, but there is no significant difference in a Lomb-Scargle periodogram \citep{1976Ap&SS..39..447L,1982ApJ...263..835S}.
However, the variability changes on longer timescales. It has a consistent phase from 2014 to 2024, but before that, in 2013, the amplitude vanishes or at least becomes very small. Similarly, in the ASAS-3 data, periodicity can be clearly seen in some years (e.g.\ 2003 and 2006, though with a different phase than the ASAS-SN data) but it is weak or absent at other times (e.g.\ 2005 and 2008).

\subsubsection{Color changes and spot coverage}
\label{sect:color}
\begin{figure}
    \centering
    \includegraphics[width=.45\textwidth]{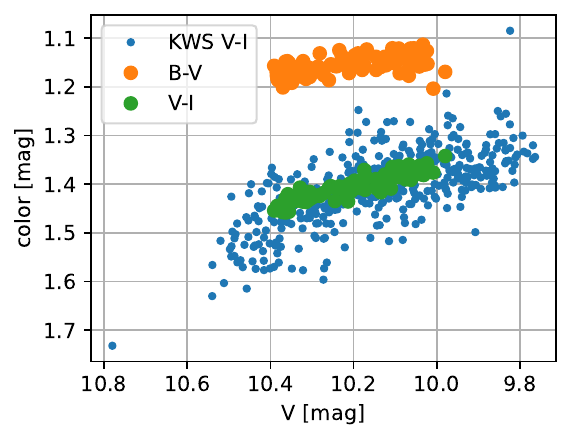}
    \caption{Color evolution of HD~251108 over several of the 21.3~d cycles (orange and green big symbols) and over years (blue small symbols).
    \label{fig:colors}}
\end{figure}

Figure~\ref{fig:colors} shows the change in color over the 21.3~d cycle, using AAVSO optical data from a single instrument and restricted to data with uncertainties below 0.01~mag. The star becomes redder when it is fainter. At its brightest, we see $V=10.0$~mag with $B-V=1.12$~mag and $V-I=1.35$~mag. The same trend it seen, albeit with larger scatter due to larger observational uncertainties, in the KWS data for $V$ magnitudes between 10.5 and 9.8.

\subsubsection{Long-term variability}
Additionally, the lightcurve shows a longer-term variability from DASCH  (Figure~\ref{fig:optlc}, top) all the way up to the most recent data.
%In the ASAS-SN data we approximated the long-term variability with a second sinusoid for the few years covered, but the longer-term DASCH, ASAS-3, and KWS data show that there is no real long-term periodicity.
The DASCH data shows brighter periods around 1900 and 1950 and a minimum in the 1920s and ASAS-3 and KWS indicate a gradual brightening between 2003 and 2012 from $m_V\approx10.0$ to $m_V\approx9.7$ followed by a slow decline to $m_V\approx10.3$ in 2021. Then, that star becomes brighter again and reaches $m_V\approx10.0$ in 2024. t is not uncommon for active stars to have more than one cycle \citep{2009A&A...501..703O,2016A&A...590A.133O}, but for our target it is not clear if the long-term variability seen in DASCH on a 100 year scale and KWS on a 20 year scale is periodic or stochastic.

\subsubsection{A spot model}

Although HD~251108 is located close to the galactic plain, the reddening is only about $E(\mathrm{B}-\mathrm{V})=0.1$~mag \citep{2019ApJ...887...93G}, much smaller than the color variations and comparable to the systematic uncertainties. Thus, we ignore this effect in the following.
We can describe the bright state with a \citet{2003IAUS..210P.A20C} spectrum with $T_\mathrm{eff}=4450\pm50$~K using the spectral libraries, tabulated filter curves, and interpolation routines from the Python packages synphot \citep{2018ascl.soft11001S} and stsynphot \citep{2020ascl.soft10003S}. If we assume that the variability is caused by star spots of a constant temperature, we can construct a spectral model that replaces some fraction of the stellar disk with a model that is cooler than $T_\mathrm{eff}=4450$. At the faint end,  $V=10.4$~mag with $B-V=1.19$~mag and $V-I=1.47$~mag.
We construct a grid of \citet{2003IAUS..210P.A20C} models with different temperature and spot covering factors and find that the faint state can be described with a model with $T_\mathrm{eff}=4100\pm100$~K and a spot covering factor of $0.6\pm0.1$, where we estimate the uncertainties in the following way: The models predict curves in covering fraction -- $T_\mathrm{eff}$ space. For a perfect fit, the three curves for $V$, $B-V$, and $V-I$ would meet in single point, but due to uncertainties in the data and our simplified model assumptions that is not the case. We thus estimate the uncertainties on covering fraction and $T_\mathrm{eff}$ by looking at the distance between the intersection points of the three curves.

In the spot model, spots seem to last over several years, as indicated by the consistent period over all of the ASAS-SN data. In the ASAS-3 and KWS, we see the periodicy disappear and reappear with a different phase. If HD~251108 is a binary, the spot does not face the companion consistently and it thus seems likely that the activity is not due to interaction with any possible wide-binary companion. The long-term changes could be related to a larger spot coverage in a location that does not rotate into and out of sight; for example large polar spots that persist for years to decades could cause the long-term changes in the lightcurve.

\subsubsection{Flare statistics}
\begin{figure}
    \centering
    \includegraphics[width=0.45\textwidth]{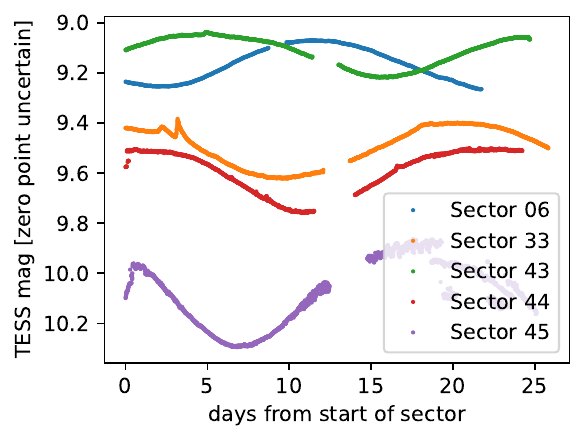}
    \caption{Five sectors of TESS data for HD~251108. Sector 45 and to a lesser degree sector 44 are contaminated by noise due to an artifact in the image subtraction. Since we use difference imaging, the absolute magnitude might be wrong, but the relative changes within an observing sector are reliable.
    \label{fig:tess_lc}}
\end{figure}

\begin{figure}
    \centering
    \includegraphics[width=0.45\textwidth]{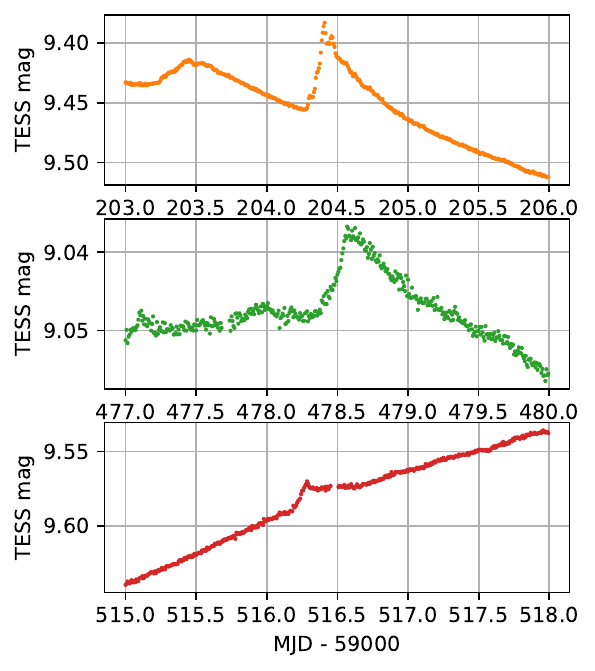}
    \caption{Visually selected flares in the TESS data.
    \label{fig:tess_flares}}
\end{figure}
Figure~\ref{fig:optlc} (middle panel) shows the flare on 2022-11-07 that is the focus of section~\ref{sect:flare} in the ASAS-SN lightcurve. This is the brightest flare observed over five observing seasons with a maximum brightening in $g$-band of 0.6~mag compared to the sinusoidal model. A second flare is seen on 2020-09-13 with a brightening of 0.4~mag.
However, the data is so sparse, that we cannot determine the peak flux and duration of those flares from the ASAS-SN data alone.

Each of the five sectors of TESS data covers about one 21.3~d period of HD~251108 (figure~\ref{fig:tess_lc}). Problems in the image subtraction lead to increased noise in sectors 44 and 45, in particular in the second half. A few small flares are visible in the TESS data by eye, but in all cases, the variability is dominated by the 21.3~d period. The shape of the lightcurve changes from cycle to cycle. Thus, we cannot reliably define the non-flaring shape of the lightcurve, which prevents us from detecting flares that are small or have complex shapes. Figure~\ref{fig:tess_flares} shows visually identified flares with a fast rise and exponential decay. Roughly, the brightening is 0.02~mag and 0.09~mag in the top panel, 0.01~mag in the middle panel, and 0.02~mag in the bottom panel. Recall from section~\ref{sect:longtermphotometry} that we use image subtraction for the TESS data, and the absolute magnitude in the figure is not reliable. The $g$ band magnitudes at the time of the flare peaks are 10.8, 10.8, 10.7, and 10.8, based on the double-sinusoidal model fit to the ASAS-SN data.
Surprisingly, the flare with the largest relative change (second flare in the top panel) and the smallest (middle panel) both last for approximately one day, while the other two are shorter. All four flares occur within 5~days of the peak of the 21.3~day period.

\subsection{X-ray flux and spectra}
\label{sect:eRASSresults}
All X-ray spectra here and in section~\ref{sect:results} are modelled using an optically thin, collisionally ionized plasma model \citep[APEC,][]{2012ApJ...756..128F}. Absorption is described by the model from \citet{2000ApJ...542..914W}. We use the solar abundances from \citet{2009ARA&A..47..481A} as base reference. Fit uncertainties are always given to the 90\% level.

For spectral analysis we combine all available eROSITA data to characterize the average, quasi-quiescent X-ray properties of HD~251108 (Table~\ref{tab:xrayfit}).
%Derived model parameter TBAbs(3xVAPEC), aspl. abund and errors (90\%, cr: 2.706): are nH: 5.0 (+1.8/-1.6) [e20/cm*+2], kT1: 0.22 (+/- 0.05) [keV], EM1: 1.37 (+0.75/-0.69) [e54/cm**3]; kT2: 0.98 (+0.15/-0.12), EM2 1.15 (+0.49/-0.23); kT3: 4.19 (+1.10/-0.75) EM3: 15.96 (+1.25/-1.28); Chi**2 is 107.1 (107 dof). Using 505 pc distance, observed/emitted Lx is 2.2/2.6 e32 erg/s (0.2-5.0 keV).
While a good fit is already obtained with solar abundances, there is residual emission around 0.5~keV (\ion{N}{7}) and a free N abundance gives N= 10.1 (+10.8/-5.5). Active stars often show an abundance pattern called IFIP (inverse first ionization potential), where elements with a high FIP (e.g.\ Ne) are enhanced over those with a low FIP (e.g.\ Fe). For this spectrum O/Fe = 4.4 is the best fit value. Overall, we find an observed (emitted) X-ray luminosity of $L_\mathrm{X} = 2.2~ (2.7) \times 10^{32}$ erg\,s$^{-1}$ in the 0.2\,--\,5.0 keV energy range. The derived X-ray properties show a highly active star with an activity level of $\log L_{\rm X}/L_{\rm bol} \approx -3$ (where $L_\mathrm{bol}$ is calculated from the bright-state $V$ band magnitude in Figure~\ref{fig:optlc}, the GAIA distance given in section~\ref{sect:target}, and the bolometric correction from \citet{1996ApJ...469..355F}, ignoring interstellar reddening and stellar variability). The coronal emission of HD~251108 is thus at saturation level of magnetic activity \citep{2011ApJ...743...48W}. The very high plasma temperatures of 30\,--\,50~MK complete this picture.
The $\log L_{\rm X}$ in eRASS is fully consistent with the earlier ROSAT and XMM-Newton data.

%JR comments: 1. only moderate improvement in fit quality and large error for O/Fe, Ne etc. unconstrained. Similar results with bin3 spectrum using cstat and 260 dof.

\subsection{Discussion}
Consistent with its location as an evolved star with detectable Li, the optical lightcurves of HD~251108 indicate a star with considerable level of stellar activity over the course of at least the last century. Several resolved flares in TESS show that this activity is magnetic, and the changes in color and brightness can be explained by large, cool spots on the stellar surface.
The quasi-quiescent X-ray properties match expectations for a highly active star with plasma temperatures of 30\,--\,50~MK and an activity level of $\log L_{\rm X}/L_{\rm bol} \approx -3$, i.e.\ at saturation level of magnetic activity.

\citet{2005A&A...444..531G} observed a correlation of X-ray flux with rotation period on single G giants, which they also interpreted as a larger surface fraction covered with active regions. The K-type giant discussed here is more active even in quiescence than G-type giants with similar rotation rate in that sample, probably because a deeper convention zone can develop \citep{2000A&A...361..614P}. Thus, it is not surprising that large X-ray flares might also happen on HD~251108.

%This increases the likelihood of magnetic field lines interacting and thus leads to larger and more frequent flares; in that picture, we would not expect such a large flare on a relatively slow rotator.

%We combine data from different telescopes to determine properties of the bright flare and its origin. We are confident to assign HD~251108 as the source of all observed activity.

\section{The large flare on HD~251108}
\label{sect:flare}

\subsection{Observations} \label{sect:data}
In addition to the ASAS-SN data discussed in section~\ref{sect:asassn} and the LEIA data from \citet{2022ATel15748....1L} we obtained data from several optical and X-ray telescopes.

\subsubsection{Optical spectroscopy}
\label{sect:optspec}
%Spectroscopic follow-up observations started shortly after the flare discovery by NICER.
We obtained a series of optical spectra with the TIGRE facility, which is located at the La Luz Observatory
near Guanajuato in central Mexico. TIGRE is a fully robotic
spectroscopy telescope which provides high-resolution spectroscopy
with a spectral resolution of a little over 20000 from 3800\,\AA{} to 8800\,\AA, thus covering important chromospheric
lines of hydrogen (H$\alpha$ and the Balmer series), calcium (Ca II H \& K and
the infrared triplet) and helium; a detailed description of TIGRE is provided by \cite{2014AN....335..787S} and \cite{2022FrASS...9.2546G}.
We also obtained spectroscopic follow-up data with the fiber-fed integral field
spectrograph \citep[KOOLS-IFU,][]{2019PASJ...71..102M} mounted on the 3.8~m Seimei telescope \citep{2020PASJ...72...48K}
at the Okayama observatory in Japan and measured H$\alpha$ fluxes.
These observations used the VPH683 grism with a wavelength coverage of $5800$ - $8000$ \AA , and
a spectral resolution of $\lambda / \Delta \lambda \sim 2000$. We also use $r$-band data from \citet{2022ATel15775....1X} taken by the BOOTES-4MET telescope at the Lijiang Observatory of the Yunnan Observatories of China \citep{2020ApJS..247...49X}.

\subsubsection{Swift}
\swift is an observatory specialized on the X-ray follow-up of gamma ray bursts, but is also used to monitor other transient sources. Its X-ray telescope \citep[XRT,][]{2005SSRv..120..165B} covers the bandpass from 0.2 to 10~keV and can detect stellar flares. \swift observed HD~251108 over five ObsIDs to follow-up the flare. An initial analysis is given by \citet{2022ATel15754....1L}. Here, we make use of the first four ObsIDs which are taken within 10~days of the flare peak.  Figure~\ref{fig:lc_NICER} has data behind the figure available electronically, where ObsIDs and observations details are listed. We retrieved fully reduced spectra and associated data products from the UK Swift Science Data Center at the University of Leicester \citep{2007A&A...469..379E}. \swift confirms that the source of the flare is compatible with HD~251108 with a positional uncertainty of 3.5 arcsec \citep{2022ATel15754....1L}. Therefore, we assume in the following that both the pre-flare X-ray emission and the large flare originate on HD~251108.

\subsubsection{NICER}
\nicer is a non-imaging instrument with a field of view of about 3~arcmin in radius. We know from ROSAT, XMM-Newton, and eROSITA that no other persistent X-ray source is present in that area so we confidently assign the X-ray emission detected by \nicer to HD~251108.

We started \nicer analysis with the raw, level-1 data publicly available on HEASARC archive. These data were reduced using the standard {\it nicerl2} task from HEASOFT V6.29c with the default screening criteria. For more details on the data reduction please see \citet{2022NatAs...6..249P}. We used the 3c50 model to estimate the empirical background spectra \citep{2022AJ....163..130R}. To extract time-resolved X-ray spectra we used the exact methodology outlined in \citet{2023NatAs...7...88P}. Figure~\ref{fig:lc_NICER} has data behind the figure available electronically, which details of the time segments used in this study.

\subsection{Results}
\label{sect:results}

%All X-ray spectra are modelled using an optically thin, collisionally ionized plasma model \citep[APEC,][]{2012ApJ...756..128F}. Absorption is described by the model from \citet{2000ApJ...542..914W}. We use the solar abundances from \citet{2009ARA&A..47..481A} as base reference. Fit uncertainties are always given to the 90\% level.

%We continue discussing the light curve (section~\ref{sect:lightcurve}), the X-ray abundances (section~\ref{sect:xrayabund}), and the X-ray spectra before (section~\ref{sect:eRASSresults}) and during the flare (section~\ref{xray:spectra}).

\subsubsection{Flare light curves}
\label{sect:lightcurve}
\label{sect:flare_lightcurve}

Figure~\ref{fig:lc_NICER} shows the various lightcurves around the flare with the top panel showing the X-ray data.
LEIA detected the flare in nine observations, resolving the flare rise phase and the peak. The data is discussed in \citet{2022ATel15748....1L} and \citet{2023arXiv230514895L}, who give two measured flux values which we show in figure~\ref{fig:lc_NICER} (top panel); the flux peaks around MJD 59891.5 \citep{2022ATel15748....1L}.

The NICER observations started when the flare was already in an exponentially declining phase.
The decay time of the flare in X-rays is an important indicator of the flare properties. The first part of the flare light curve measured in NICER before about MJD 59901 can be described well with an exponential decay with a decay time of 2.2~days on top of a constant flux around $\log L_\mathrm{X} = 32.7$ (in $\mathrm{erg}\;\mathrm{s}^{-1}$). That fitted constant flux is several times larger than the X-ray flux seen before the flare. Extrapolating the decay fitted to the NICER data to the earlier LEIA observations shows that it falls short by about a factor of two below the observed peak flux.
After MJD 59901, we see a short-lived marginal increase in X-ray flux, followed by a steeper decay (second panel in figure~\ref{fig:lc_NICER}) until the X-ray flux reaches the quiescent level after about 25~days.

The third panel of Fig.~\ref{fig:lc_NICER} shows the $g$ band flux from the ASAS-SN survey which caught the optical flare on MJD 59890.9 with the $g$-band flux increasing from about 200~mJy to 380~mJy. The next ASAS-SN observation occurred three days later and is no longer enhanced over the sinusoidal variability pattern.

\begin{figure}
    \centering
    \includegraphics[width=0.45\textwidth]{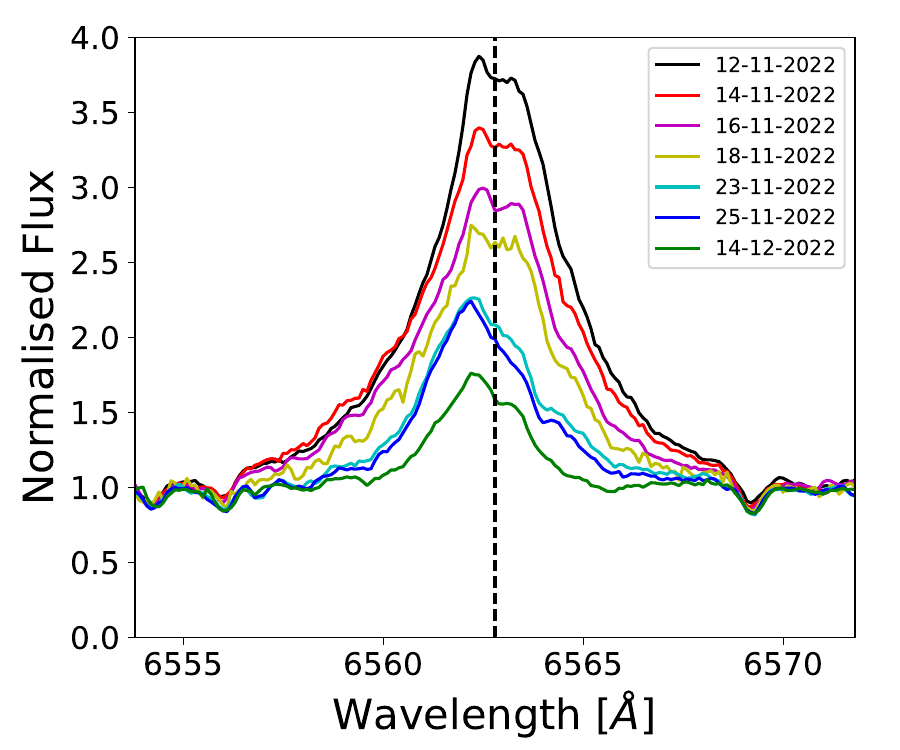}
    \caption{Examples of the H$\alpha$ line profile observed with TIGRE.
    \label{fig:lc_Ha_TIGRE}}
\end{figure}

\begin{figure*}
    \centering
    \includegraphics[width=0.8\textwidth]{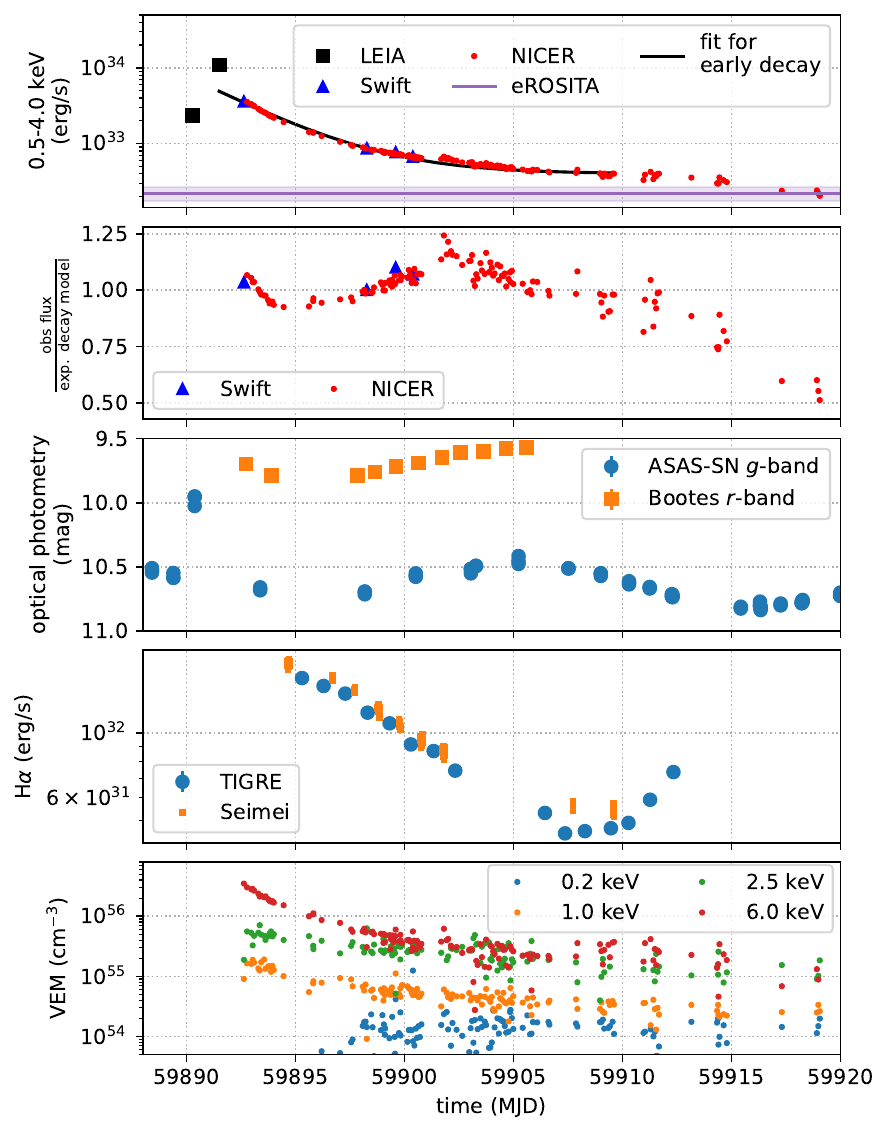}
    \caption{Lightcurve for HD~251108. Where error bars are invisible, the statistical uncertainties are smaller than the plot symbols. \emph{top:} Absorbed X-ray flux in the 0.5-4.0~keV range. The average flux in eRASS is marked by a purple line, the observed variability (shaded) is also consistent with the XMM-Newton and ROSAT fluxes.
    The black line shows the best-fit exponential decay fitted to the early decay phase, see section~\ref{sect:lightcurve} for details.
    \emph{second panel:} Ration between X-ray flux and the simple exponential decay model.
    \emph{third panel:} $g$-band lightcurve from ASAS-SN. The survey uses different telescopes and there are small systematic offsets between individual cameras, which cause almost simultaneous measurements to differ slightly. \emph{fourth panel:} Equivalent width of the H$\alpha$ emission line. \emph{bottom:} Volume emission measure of the flare from a fit with four components at fixed temperature and abundance. See section~\ref{xray:spectra} for details. The hotter flare components decay faster, while the 0.2~keV component is mostly stable. Error bars are omitted for clarity. A full listing of all NICER observations, all the data in this figure \citep[except for the 2 datapoints from LEIA see][]{2022ATel15748....1L}.% and in Figure~\protect{\ref{fig:logTlogEM}} is available electronically as ``data behind the figure''.
    \label{fig:lc_NICER}}
\end{figure*}

The fourth panel in Fig.~\ref{fig:lc_NICER} shows the equivalent width (EW) of the H$\alpha$ emission line.
There are small systematic differences between instruments due to their different spectral resolution but both show the same trend.
Monitoring in H$\alpha$ starts about three days after the X-ray peak and the equivalent width declines linearly for about 10 days, then begins to rise again. Given the better spectral resolution of TIGRE, we use that dataset for fitting and find an exponential decay time of 9.8~days in the same time interval MJD 59891.5 to 59901 used to fit the X-ray decay. After MJD 59908, the H$\alpha$ equivalent width increases again without an associated increase in X-ray activity. This increase may be due to some active regions with spots and plages, but without associated X-ray emitting structures, rotating into view; those regions would be cooler, consistent with the decreasing flux in $g$ at the same time. Figure~\ref{fig:lc_Ha_TIGRE} shows a few H$\alpha$ line profiles from TIGRE; there is little variation in the line profile, but the equivalent width changes significantly.

The bottom panel shows the emission measure evolution from the NICER data, discussed in more detail in the section~\ref{xray:spectra}.

\subsubsection{X-ray abundances}
\label{sect:xrayabund}
We grouped the NICER data into different periods (MJD 59892-59895, 59895-59901, 59901-59907) and determined the abundances for each of them by fitting a model with three emission components and one absorption component. For this fit, temperature and emission measure in each component are allowed to float independently to find the best description of the thermal structure of the emitting plasma. Because the spectrum does not contain line-free regions but is dominated by a pseudo-continuum of unresolved spectral lines, absolute abundance measurements are always ambiguous. Thus, we keep the oxygen abundance fixed at 1 relative to our abundance baseline from \citet{2009ARA&A..47..481A}, and vary other abundances relative to that. We find that the absorbing column density and abundances measured agree for the three different time periods within the errors; there is no evidence that the coronal abundance changes during the flare. Since the relative contribution of the background is the lowest at early times, when the count rate is highest, we chose the fit obtained during that period as canonical set of abundances that we will use for the remainder of the analysis; that fit is shown in table~\ref{tab:xrayfit}.
% C, N, O, Ne relative to Fe
% Asplund 2009 has
%6   8.43  C
%7   7.83  N
%8   8.69  O
%10   7.93  Ne
%26   7.50  Fe
% https://ui.adsabs.harvard.edu/abs/2022A%26A...659A...3S/abstract uses
% A(H) = 12: A(C) = 8.43, A(N) = 7.83, A(O) = 8.69, A(Ne) = 8.30
% so, we need to correct our Ne values by 0.43
% with that, our canonical abudances are
% mean([1,1,1,2.3 * 0.43]) / 0.33 ~ 3
% np.log10(3) + 0.048 -> 0.53

\citet{2022A&A...659A...3S} looked at the difference of photospheric and coronal abundances depending on the first ionization potential (FIP) of the elements. Without detailed measurements, they assume solar photospheric abundances from \citet{2009ARA&A..47..481A} \citep[except for Ne, which is taken from ][]{2005Natur.436..525D} for the photosphere. For each of C, N, O, and Ne, they calculate the logarithm of the ratio with respect to Fe. Our data is insufficient to use the pre-flare abundances, but since those support an IFIP abundance pattern, just like we observe in the flare, and we do not find any change in the abundance pattern from the flare peak to the decay, it seems reasonable to assume that the abundances measured in the flare represent the coronal abundance of HD~251108. Just like O, N and C are fixed at the photospheric value in our fits, and the Ne abundance is very close to the number from \citet{2005Natur.436..525D}. So, the value we measure in the flare for the average logarithm of the ratio is [X/Fe]=0.48; following \citet{2022A&A...659A...3S} we correct that number by +0.048 \citep{2018ApJ...862...66W} and end up with [X/Fe]=0.53. This places HD~251108 in the upper branch of the FIP-$T_\mathrm{eff}$ diagram.

\begin{table*}
    \caption{X-ray fits \label{tab:xrayfit}}
    \centering
    \begin{tabular}{cccccc}
        \hline\hline
        Parameter & unit & eROSITA & 59892-59895 & 59895-59901 & 59901-59912\\
        \hline
        $N_\mathrm{H}$ & $10^{20}$ cm$^{-2}$ & $5.0^{+1.8}_{-1.6}$ & $5.31\pm 0.07$ & $5.2^{+0.3}_{-0.5}$  & $4.8^{+1.3}_{-0.4}$\\
        k$T_1$ & keV & ${0.22 \pm 0.05}$ & $0.86^{+0.11}_{-0.03}$ &  $0.81^{+0.15}_{-0.07}$ & $0.97^{+0.04}_{-0.26}$\\
        k$T_2$ & keV & $0.98^{+0.15}_{-0.12}$ & $1.7^{+2.8}_{-0.2}$ & \tablenotemark{b} & \tablenotemark{b}\\
        k$T_3$ & keV & $4.2^{+1.1}_{-0.8}$ & $6.^{+0.5}_{-0.3}$ & $3.8^{+1.3}_{-0.1}$ & $3.9^{+0.6}_{-3.2}$ \\
        $VEM_1$ & $10^{54}$ cm$^{-3}$ & $1.4^{+0.8}_{-0.7}$& $11.0^{+1.1}_{-0.9}$ &$5.^{+4.}_{-1.}$ & $9.3^{+0.5}_{-6.}$ \\
        $VEM_2$ & $10^{54}$ cm$^{-3}$ & $1.2^{+0.5}_{-0.2}$ & $28.^{+200.}_{-8.}$ & \tablenotemark{b} & \tablenotemark{b} \\
        $VEM_3$ & $10^{54}$ cm$^{-3}$ & $16.0\pm{1.3}$ & $243.^{+8.}_{-207.}$ & $80.2^{+0.7}_{-16.6}$ & $50.^{+6.}_{-2.}$\\
        O & \tablenotemark{a} & =1 & =1 & =1 & =1\\
        Ne &\tablenotemark{a} & =1 & $2.3^{+1.4}_{-0.3}$  & $2.7^{+0.6}_{-1.2}$ & $1.2^{+2.0}_{-0.5}$ \\
        Mg &\tablenotemark{a} & =1 & $0.65\pm 0.09$ & $0.3^{+0.5}_{-0.3}$ & $0.4^{+0.6}_{-0.4}$ \\
        Si &\tablenotemark{a} & =1 & $0.5\pm 0.07$ & $0.2^{+0.4}_{-0.2}$ & $0.3^{+0.4}_{-0.3}$\\
        S &\tablenotemark{a} & =1 & $0.2\pm 0.1$ & $0.1^{+0.6}_{-0.1}$ & $0.04^{+0.62}_{-0.04}$\\
        Fe &\tablenotemark{a} & =1 & $0.33^{+0.12}_{-0.03}$ & $0.25^{+0.06}_{-0.08}$ & $0.16^{+0.15}_{-0.05}$\\
        Ni &\tablenotemark{a} & =1 & $1.5^{+0.3}_{-0.9}$ & $1.\pm 1.$  & $0.8^{+2.4}_{-0.8}$ \\
        dof & & 107 & 779 & 731 & 709\\
        red. $\chi^2$ & & 1.0 & 1.1 & 1.0 & 1.0\\
        \hline
    \end{tabular}
    \tablenotetext{a}{Relative to \citet{2009ARA&A..47..481A}}
    \tablenotetext{b}{Within the errors, the emission measure of this component could be zero, and thus the temperature is not constrained; in other words, this component is not required for the fit.}
\end{table*}

\subsubsection{X-ray spectra}
\label{xray:spectra}
\begin{figure}
    \centering
    \includegraphics[width=0.45\textwidth]{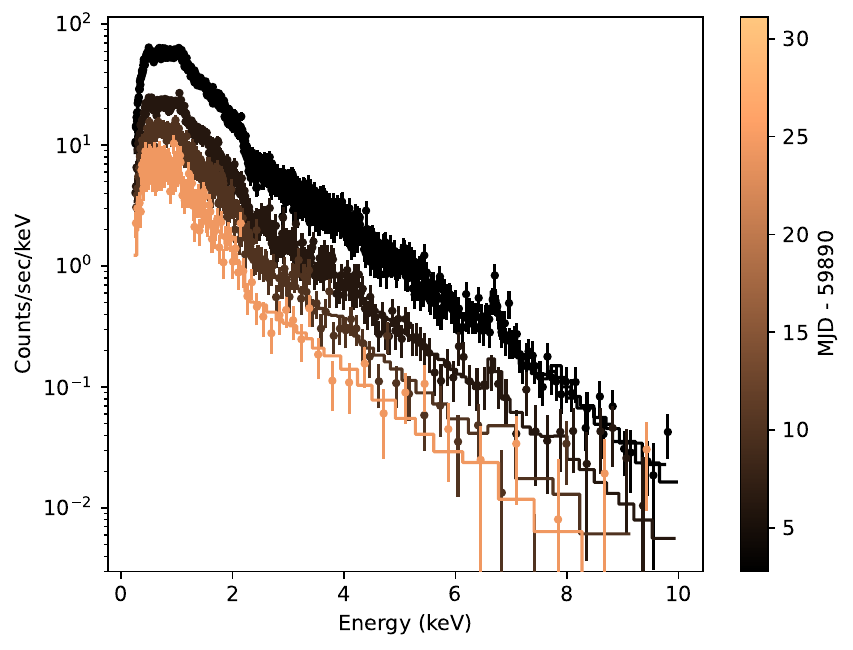}
    \caption{Evolution of NICER spectra during the flare; shown are four spectra from one orbit each. Spectra are binned to a minimum of 25 counts per bin. Colors correspond the time of the observation.
    \label{fig:NICER_spectra}}
\end{figure}

Figure~\ref{fig:NICER_spectra} shows NICER spectra observed at different periods. The flux decreases over time and in particular the Fe 6.7~keV line is no longer visible in later spectra. The figure also gives a good indication of the data quality - as the flare decays, the count rate, and thus the signal in each individual observation, decreases, in particular at high energies. As the count rate decreases, the relative importance of the background increases, leading to a few erroneously high bins $>8$~keV in the later spectra.

After fixing the abundances (section~\ref{sect:xrayabund}), the \nicer spectra can be fit well with just a few parameters. We can either use three temperature components and fit temperature and emission measure (six parameters) or pick a fixed temperature grid  and just vary the emission measure (four parameters when choosing four temperatures). In the first approach, we consistently find emission components around 0.2 and 1.0~keV, as also seen in the eROSITA observations, and a hotter component that declines in temperature.
The $\chi^2$ indicates that the spectra may be overfit with six parameters; on the other hand, fits with just two temperature components systematically underpredict spectral bins at higher energies. We thus prefer a description with a fit of four components at fixed temperature. Figure~\ref{fig:lc_NICER} (bottom panel) shows the evolution the emission measure for those fixed temperatures.
The four \swift spectra have much fewer counts than the \nicer spectra. However, for consistency, we use the same model with four fixed temperature to fit them, even though the $\chi^2$ shows that the model is overfitting the data. The fitted values for the emission measure however are consistent with the \nicer data.

The hottest component (6.0~keV) dominates the emission early in the flare. Figure~\ref{fig:lc_NICER} (bottom panel) shows an apparent absence of 0.2~keV plasma before MJD 59897, but we caution that this is likely a fitting artifact: The extreme contrast between the hot and cool components means that the cool components are lost in the statistical noise of the much brighter hot components and the fit of the cool components is not well constrained.
The 1.0~keV component also decays fast, while the 2.5~keV component is more persistent; possibly because it picks up plasma from the 6.0~keV component that cools down over time.

\subsection{Discussion}
\label{sect:discussion}

We discuss an extraordinarily long and energetic flare observed in X-ray and optical data with a peak flux around $10^{34}$~erg~s$^{-1}$ (0.5-4.0 keV). The integrated soft X-ray flux is about five orders of magnitude stronger than the Carrington event \citep[$L_X = 10^{28}$~erg~s$^{-1}$ in the 1-8~\AA{} band (in that band the flare discussed here would show about $10^{33}$~erg~s$^{-1}$),][]{2013JSWSC...3A..31C}, one of the strongest known solar space weather events. Of course, young stars are significantly more active than the Sun, but the peak X-ray flux in the HD~251108 flare is still an order of magnitude above the strongest flare observed in the Chandra Orion Ultradeep Project which monitored over 1400 pre-main sequence stars for almost 1 Ms and observed hundreds of flares \citep{2008ApJ...688..418G}. Since giant stars are more distributed over the sky, there is less monitoring, but \citet{2023MNRAS.518..900K} compiled all known large flares from giant stars. RS CVn star \object{GT Mus} monitored by \citet{2021ApJ...910...25S} shows several flares with a longer decay time, but none reach the peak X-ray flux observed in HD~251108.

In Section~\ref{sect:flare_lightcurve} we fit a decay time of 2.2~days for the early phase of NICER observations, but noted that the LEIA data of the flare peak is about twice as bright as an extrapolation of this trend. Also, at later times, the NICER curve begins to drop faster. Taken together, this shows that the flare evolution is more complex than a simple exponential decay as often modeled. Still, to obtain some estimate of the dimensions of the flaring loop, we will apply such a model to the NICER date before MJD 59891.5. Integrating just the decaying exponential in Fig.~\ref{fig:lc_NICER} over this time range gives a flare energy around $1\times 10^{39}$~erg in the 0.5-4.0~keV band. The fitted absorbing column density is so small that the intrinsic flux is only about 7\% larger than that. For the H$\alpha$ lightcurve we find a 9.8~day exponential decay time and an integrated energy flux of $1.2\times 10^{38}$~erg in the H$\alpha$ line. \citet{2022PASJ...74..477K} study the relation between integrated flare energy in X-rays $E_X$ and H$\alpha$  $E_{\mathrm{H}\alpha}$ and find the X-ray energy to be ten times larger than H$\alpha$ energy for a wide range of X-ray flares with $E_X$ from $1\times 10^{29}$~erg to $10^{38}$~erg using the wide X-ray band from 0.1-100~keV. We extrapolate the spectral models from section~\ref{xray:spectra} to that band, and find a corrected flare flux of around $E_\mathrm{X_\mathrm{0.1-100 keV}} = 1.5\times 10^{39}$~erg; the flare described here thus matches the  $E_X/E_{\mathrm{H}\alpha}$-ratio observed in the sample of fainter flares from \citet{2022PASJ...74..477K}. Our new observations extend that relation by one order of magnitude. This indicates that the physical relation that connects the X-ray emission, originating from the hot plasma in the flare loop, and the H$\alpha$ emission, originating closer to the stellar surface, still holds for flares as energetic as the one described in this work.

\begin{figure}
    \centering
    \includegraphics[width=0.45\textwidth]{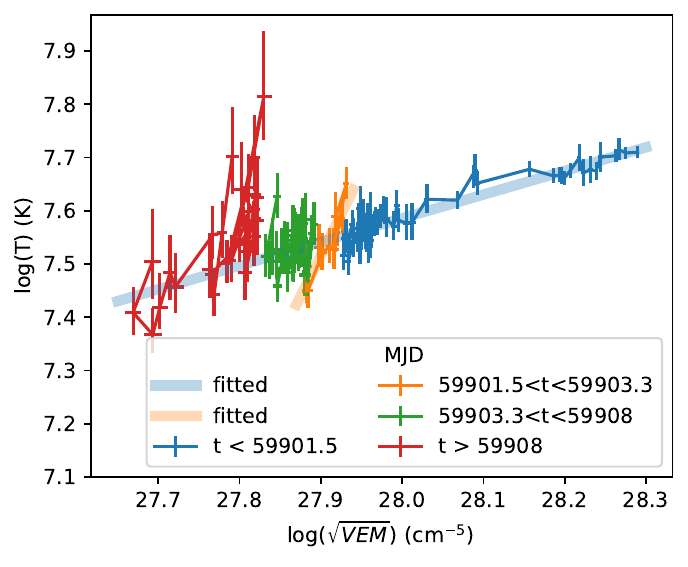}
    \caption{The $0.5 \log \sqrt{VEM} - \log T$ diagram for a single temperature model. The thick, partially transparent blue line is the best fit to the data before MJD=59901.5 and also describes the data after MJD 59903.3 reasonably well. The thick, partially transparent orange line is the best fit for the period in between. It is much steeper, consistent with a little or no re-heating. Data shown in this figure is included in the electronic ``data behind the Figure'' for figure~\ref{fig:lc_NICER}.
    \label{fig:logTlogEM}}
\end{figure}

The hot plasma in the flare loop cools by radiation and conduction; the relative importance of both processes depends on the temperature, density, and geometry of the flare loop. At the same time, it is possible that the plasma is continuously re-heated through shock waves or further magnetic activity. Through combination of analytic approximations, scaling laws fitted to numerical simulations, and certain assumptions (e.g.\ a common assumption is that the length of a loop is ten times its cross-section), one can derive formulas that allow an estimate of loop length, density, and re-heating from observed quantities such as the observed peak X-ray temperature, the exponential decay time of the X-ray flux, and the slope of the X-ray flare decay in a $\log \sqrt{VEM} - \log T $ diagram \citep[see][and references therein for a review and discussion of these formulas]{2023MNRAS.518..900K}. For most X-ray flares described in the literature, the signal in the decay is not sufficient to fit and track individual temperature components as we have done in figure~\ref{fig:lc_NICER}. Instead, \citet{2007A&A...471..271R} developed scaling factors to relate the temperature of a one temperature model to the broad distribution of temperatures seen in real flares and hydrodynamical models. Those factors depend on the band-pass and response of the specific instrument and need to be calibrated with models. Even though our data is sufficient to fit a model with multiple components and non-solar abundances, we derive the loop length in this framework for comparison to other observations of superflares in the literature. Values for the scaling factors for NICER have not been published, so we use the numbers given for XMM/EPIC or Chandra/ACIS in \citet{2007A&A...471..271R} because the effective area curve for these instruments has a similar shape to NICER. We repeat the fit from section~\ref{xray:spectra} with just a single temperature component and show the resulting $\log \sqrt{VEM} - \log T$ diagram in figure~\ref{fig:logTlogEM}.

For the early phase of the flare (blue), we find a slope of $\zeta = 0.44\pm0.02$ corresponding to significant ongoing re-heating \citep{1993A&A...267..586S,2014LRSP...11....4R}. With this slope, the decay time of 2.2~days, and an extrapolation of the temperature of a single temperature plasma observed with NICER to the peak of the flare, the scaling from \citet{2007A&A...471..271R}, gives a loop length of $L = 12\;R_\sun{} \approx 1.5-2\;R_*$. The formula from \citet{2023MNRAS.518..900K} leads to an estimate around $L = 22\;R_\sun{}\approx2.8-3.7\;R_*$ with similar assumptions as does appendix \ref{sect:Shibata_Yokoyama}; either way, the loop semi-length is comparable to the stellar radius. This is much higher above the atmosphere than observed in super-flares on main-sequence stars like EV~Lac, where \citet{2010ApJ...721..785O} found a flare with a peak temperature of 12~keV, but a semi-loop length of only $0.4\;R_*$, see \citet{2023ApJ...944..163H} for similar observations in a solar-type star.

Around MJD 59901.5, the slope steepens considerably for about 1.8 days, consistent with a period of no re-heating (orange in figure \ref{fig:logTlogEM}), before it settles to a $\zeta$ consistent with the earlier decay again (green). This indicates a major change in the plasma conditions, likely the structure of the magnetic field, to suppress re-heating for about 1.8 days, before re-heating resumes as before. This suggests that the flare geometry is more complex than a single loop; the length estimates have to be regarded as upper limits.

We do not detect any change in abundance during the flare discussed here; this is in contrast to a large flare on Algol where \citet{1999A&A...350..900F} claim a three-fold increase in abundance. Either HD~251108 is different from Algol because it is active enough to show a strong IFIP abundance pattern in quiescence already, or the fits of Algol data by \citet{1999A&A...350..900F} are impacted by an ambiguity between abundance and temperature often seen in fits with too few temperature components - an issue we can avoid because the \nicer signal is strong enough to allow us to fit four independent temperature components.

\section{Summary}
\label{sect:summary}
HD~251108 is an evolved K-type giant in the Li-rich, active sub-group. It shows rotational modulation with a period of 21.3~days which can be explained by large stellar spots as rotate in and out of view. Those spots are stable for several years. In addition, it displays photometric variability on time scales of one or more decades of order 0.5~mag; again, this is consistent with very large and very stable stellar spots. The quasi-quiescent X-ray flux of HD~251108 is at the saturation level with $\log L_{\rm X}/L_{\rm bol} \approx -3$.

We follow the decay of a superflare on K-Giant HD~251108 with NICER until it reaches the X-ray flux observed in eRASS pre-flare about 28~days after the flare peak. We present data from about 130 epochs of NICER, each of which has sufficient signal to fit a model with multiple temperature components and non-solar abundances. In a model with four temperature components, the hotter components decay faster, while the cooler components are mostly stable. Abundances are stable throughout the flare and consistent with typical active stars with an IFIP effect as seen by an Ne/Fe ratio that is about a factor of ten larger than in the Sun. In the initial decay, the X-ray lightcurve is matched by a decay in the H$\alpha$ flux, while the plasma shows some re-heating. We estimate the length of the flare loop to be 2-4 times larger than the radius of the star. About 10 days after the flare peak, the flare undergoes a short phase of limited re-heating and the lightcurve begins to deviated from the initial decay. This is one of the strongest flares ever observed.

\begin{acknowledgements}

We want to thank the anonymous referee for their very constructive and detailed comments that helped to improve the manuscript.
This research has made use of the SIMBAD database,
operated at CDS, Strasbourg, France \citep{2000A&AS..143....9W}.
This research has made use of the VizieR catalogue access tool, CDS,
 Strasbourg, France (DOI : 10.26093/cds/vizier). The original description
 of the VizieR service was published in \citet{2000A&AS..143...23O}.
This research has made use of NASA's Astrophysics Data System Bibliographic Services.
This work has made use of data from the European Space Agency (ESA) mission
{\it Gaia} (\url{https://www.cosmos.esa.int/gaia}), processed by the {\it Gaia}
Data Processing and Analysis Consortium (DPAC,
\url{https://www.cosmos.esa.int/web/gaia/dpac/consortium}). Funding for the DPAC
has been provided by national institutions, in particular the institutions
participating in the {\it Gaia} Multilateral Agreement. This work uses data from eROSITA, the soft X-ray instrument aboard SRG. It also uses observations obtained with XMM-Newton, an ESA science mission with instruments and contributions directly funded by ESA Member States and NASA.
This work made use of data supplied by the UK Swift Science Data Centre at the
University of Leicester. We acknowledge with thanks the variable star observations from the AAVSO International Database contributed by observers worldwide and used in this research.

HMG was supported by the National Aeronautics and Space Administration through Award 80NSSC21K1903.
J.R. acknowledges support from the DLR under grant 50QR2105.
Y.N. acknowledge support from NASA ADAP award program Number 80NSSC21K0632. Y.N. was also supported by JSPS Postdoctoral Research Fellowship Program and
JSPS KAKENHI Grant Number 21J00106.
H.M. acknowledges support from the JSPS KAKENHI Grant Number 20K04032.
PCS acknowledges support from the DLR under grant 50OR2205.

\end{acknowledgements}

\appendix

\section{Calculation of loop size and magnetic field strength by Shibata \& Yokoyama formula}
\label{sect:Shibata_Yokoyama}
For comparison, we also calculate the flare loop size, the fundamental physical quantity of the flare by using equations based on the magnetic reconnection model by
\citet{2002Apj...577..422}. They present the following scaling laws for the length of a loop $L$ and the flare magnetic field  strength $B$:
\begin{equation} \label{eq:Shibata_Yokoyama_B}
\mathnormal{B} = 50 \left( \frac{VEM_{\mathrm{peak}}}{10^{48} \: \mathrm{cm}^{-3}} \right)^{-1/5} \left( \frac{\mathnormal{n_{0}}}{10^{9} \: \mathrm{cm}^{-3}} \right)^{3/10} \left( \frac{\mathnormal{T_{\mathrm{peak}}}}{10^{7} \: \mathrm{K}} \right)^{17/10}
\: \mathrm{G}
\end{equation}
\begin{equation} \label{eq:Shibata_Yokoyama_L}
\mathnormal{L} = 10^{9} \left( \frac{VEM_{\mathrm{peak}}}{10^{48} \: \mathrm{cm}^{-3}} \right)^{3/5} \left( \frac{\mathnormal{n_{0}}}{10^{9} \: \mathrm{cm}^{-3}} \right)^{-2/5} \left( \frac{\mathnormal{T_{\mathrm{peak}}}}{10^{7} \: \mathrm{K}} \right)^{-8/5}
\: \mathrm{cm}
\end{equation}
where $VEM_{\mathrm{peak}}$ is the peak volume emission measure, $T_{\mathrm{peak}}$ is the peak temperature, and $n_{0}$ is the preflare coronal density. From our X-ray spectral analysis, $kT_{\mathrm{peak}} \sim 4.9 \: \mathrm{keV}$ and $VEM_{\mathrm{peak}} \sim 5.9 \times 10^{56} \: \mathrm{cm}^{-3}$ are inferred.
$B$ and $L$ calculated by using  these results and equations (\ref{eq:Shibata_Yokoyama_B}) and (\ref{eq:Shibata_Yokoyama_L}) are shown in table \ref{tab:Shibata_Yokoyama}. Though we cannot know $n_{0}$ from our observation data, when $11 < \log n_{0} < 12$, these results are consistent with $L = 12-22 \: R_{\odot}$ shown in  Section \ref{sect:discussion}.

\begin{deluxetable*}{rrr}
    \tablecaption{Magnetic field  strength and flare loop size from equations (\ref{eq:Shibata_Yokoyama_B}) and (\ref{eq:Shibata_Yokoyama_L}) \label{tab:Shibata_Yokoyama}}
    \tablehead{\colhead{} & \colhead{$B$ (G)} & \colhead{$L$ ($R_{\odot}$)}}
    \startdata
    $n_{0}=10^{11} \: \mathrm{cm^{-3}}$ & $68$ & $26$ \\
    $n_{0}=10^{12} \: \mathrm{cm^{-3}}$ & $135$ & $10$ \\
    $n_{0}=10^{13} \: \mathrm{cm^{-3}}$ & $268$ & $4$
    \enddata
    \end{deluxetable*}

\facilities{NICER, XMM, ROSAT, ASAS-SN, eROSITA, Swift, AAVSO, TESS}

\software{AstroPy \citep{2013A&A...558A..33A,2018AJ....156..123A}, NumPy \citep{van2011numpy,harris2020array}, Matplotlib \citep{Hunter:2007}, Sherpa \citep{2007ASPC..376..543D,2021zndo...5554957B}, synphot \citep{2018ascl.soft11001S}, stsynphot \citep{2020ascl.soft10003S}}

\bibliography{bib}{}
\bibliographystyle{aasjournal}

%% This command is needed to show the entire author+affiliation list when
%% the collaboration and author truncation commands are used.  It has to
%% go at the end of the manuscript.
%\allauthors

%% Include this line if you are using the \added, \replaced, \deleted
%% commands to see a summary list of all changes at the end of the article.
%\listofchanges

\end{document}